# Design of $N_nH_{3n+1}^+$ Series of "Non-Metallic" Superalkali Cations

Ambrish Kumar Srivastava

*Department of Physics, DDU Gorakhpur University, Gorakhpur- 273009, India*

E-mail: ambrishphysics@gmail.com; aks.ddugu@gmail.com

**Abstract**

The species with lower ionization energy than alkali atoms are referred to as superalkalis. Typical superalkalis include a central electronegative core with excess metal ligands. We propose a new series of "non-metallic" $N_nH_{3n+1}^+$ superalkali cations using MP2/6-311++G(d,p) level. These cations are designed by successive replacement of H-ligands of ammonium cation ($NH_4^+$) by ammonium ($NH_4$) moieties. The resulting $N_nH_{3n+1}^+$ cations, which can be expressed in the form of $[NH_4 \cdots (n-1)NH_3]^+$ complexes, possess a number of unusual N-H⋯N type of partially covalent H-bonds, with the interacting energy in the range 7.8-24.3 kcal/mol. These cations are stable against loss of a proton ($N_nH_{3n}+H^+$) and loss of ammonia $[(n-1)NH_3+NH_4^+]$. The vertical electron affinities ($EA_v$) of $N_nH_{3n+1}^+$ decreases monotonically from 4.39 eV from $n = 1$ to 2.39 eV for $n = 5$, which suggest their superalkali nature. This can be explained on the basis of electron localization on core (central) N-atom ($Q_c$), as $EA_v$ correlates linearly with $Q_c$. We have also demonstrated that this series may be continued to obtain new superalkali cations with even lower $EA_v$, by exemplifying $N_9H_{28}^+$ with the $EA_v$ of 1.84 eV. $N_9H_{28}^+$ is stabilized by four partially covalent H-bonds (8.5 kcal/mol each) and four electrostatic H-bonds (0.4 kcal/mol each). This led to an exponential relation between $EA_v$ and $n$, which may provide an approximate $EA_v$ of for any value of $n$ in $N_nH_{3n+1}^+$ series.

## 1. Introduction

The species with lower ionization energy (IE) than alkali atoms are referred to as superalkalis. According to Gutsev and Boldyrev,[1] such species can be designed by a central electronegative atom (X) with excess electropositive ligands (M), assuming a general formula of $XM_{k+1}^+$, where $k$ is the valance of X. Typical examples of superalkali cations include $FLi_2^+$, $OLi_3^+$, $NLi_4^+$ etc. Superalkalis are hypervalent clusters possessing an excess electron and hence, strong reducing capability. Consequently, superalkalis can be employed in the formation of a variety of charge transfer species with unusual properties. For instance, their applications in the design of superbases with strong basicity,[2,3] supersalts with tailored properties,[4-6] alkalides with negatively charged alkali metals,[7,8] *etc.* have been extensively studied. Owing to interesting properties of superalkalis and their compounds, new classes of superalkalis have been continuously explored.[9-15] Tong *et al.* have reported binuclear superalkali cations $M_2Li_{2k+1}^+$, where $k$ is valence of M (= F, O, N and C)[10] as well as polynuclear superalkali cations $YLi_3^+$ using various functional groups (Y= $CO_3$, $SO_3$, $SO_4$ etc.) as central core.[11] The polynuclear superalkalis based on alkali-monocyclic (pseudo)-oxocarbon have been also reported.[12] The non-planar aromatic superalkalis have also been designed using inorganic such as $Be_3^{2-}$, $B_3^-$, $Al_4^{2-}$ etc. as well as organic $C_4H_4^{2-}$, $C_5H_5^-$ aromatic anions.[13] Giri *et al.*[14] have studied $P_7^{3-}$ Zintl core decorated with organic ligands [R = Me, $CH_2Me$, $CH(Me)_2$ and $C(Me)_3$] and explored the superalkali behavior of $P_7R_4^+$ cations. Zhao *et al.*[15] have presented a rational design of new superalkalis such as $Al_3^+$, $B_9C_3H_{12}^+$, $C_5NH_6^+$, *etc.* and demonstrated their role in the reduction of $CO_2$, which was further explored by other workers.[16,17] Recently, Sun *et al.*[18] have proposed the design of a new series of superalkali cations, $XM_2^+$ by taking typical superalkalis (M= $FLi_2$, $OLi_3$, and $NLi_4$) as ligands and (super)halogen as a core (X).

Ammonium ion ($NH_4^+$) is one of the most popular radicals, which is equally important in both inorganic as well as organic chemistry. Structurally, $NH_4^+$ appears a non-metallic analog of $NLi_4^+$, a typical superalkali cation. Hou *et al.*[19] attempted to study non-metallic superalkalis, exploring $F_2H_3^+$, $O_2H_5^+$, $N_2H_7^+$, and $C_2H_9^+$ cations. However, our literature survey reveals that $NH_4^+$ exhibits metallic properties under certain conditions.[20,21] Therefore, it is interesting to observe whether it possesses superalkali nature. More interestingly, we explore a new series of non-metallic superalkali cations, $N_nH_{3n+1}^+$, starting from $NH_4^+$ and replacing H-ligands with $NH_4$ successively. We perform a systematic investigation on the structure, bonding, and stability of $N_nH_{3n+1}^+$ series of cations as well as the evolution of superalkali nature in this series.

## 2. Computational details

The equilibrium structure and energies of $N_nH_{3n+1}^+$ cations were obtained by the geometry optimization and single-point calculations at second order Møller-Plesset perturbation theory based method (MP2)[22] and 6-311++G(d,p) basis set. The geometry optimization was followed by vibrational frequency calculations in order to ensure that the optimized structures belong to a true minimum in the potential energy surface. In order to explore the superalkali nature, we have computed vertical electron affinity ($EA_v$) of $N_nH_{3n+1}^+$ cations by the difference of the total electronic energy of equilibrium structure of cations and single-point energy of respective neutral species at cationic geometry. This provides the energy required to attach an electron to $N_nH_{3n+1}^+$ cations, which is equivalent to the energy released to extract an electron from $N_nH_{3n+1}$ species, i.e., the IE of $N_nH_{3n+1}$ species. The partial atomic charges were obtained using natural bond orbital (NBO) analysis[23] at MP2/6-311++G(d,p) level. All computations were carried out using Gaussian 09 suite of programs.[24] The quantum theory of atoms in molecules (QTAIM) analyses[25] were performed using AIMAll software package.[26]

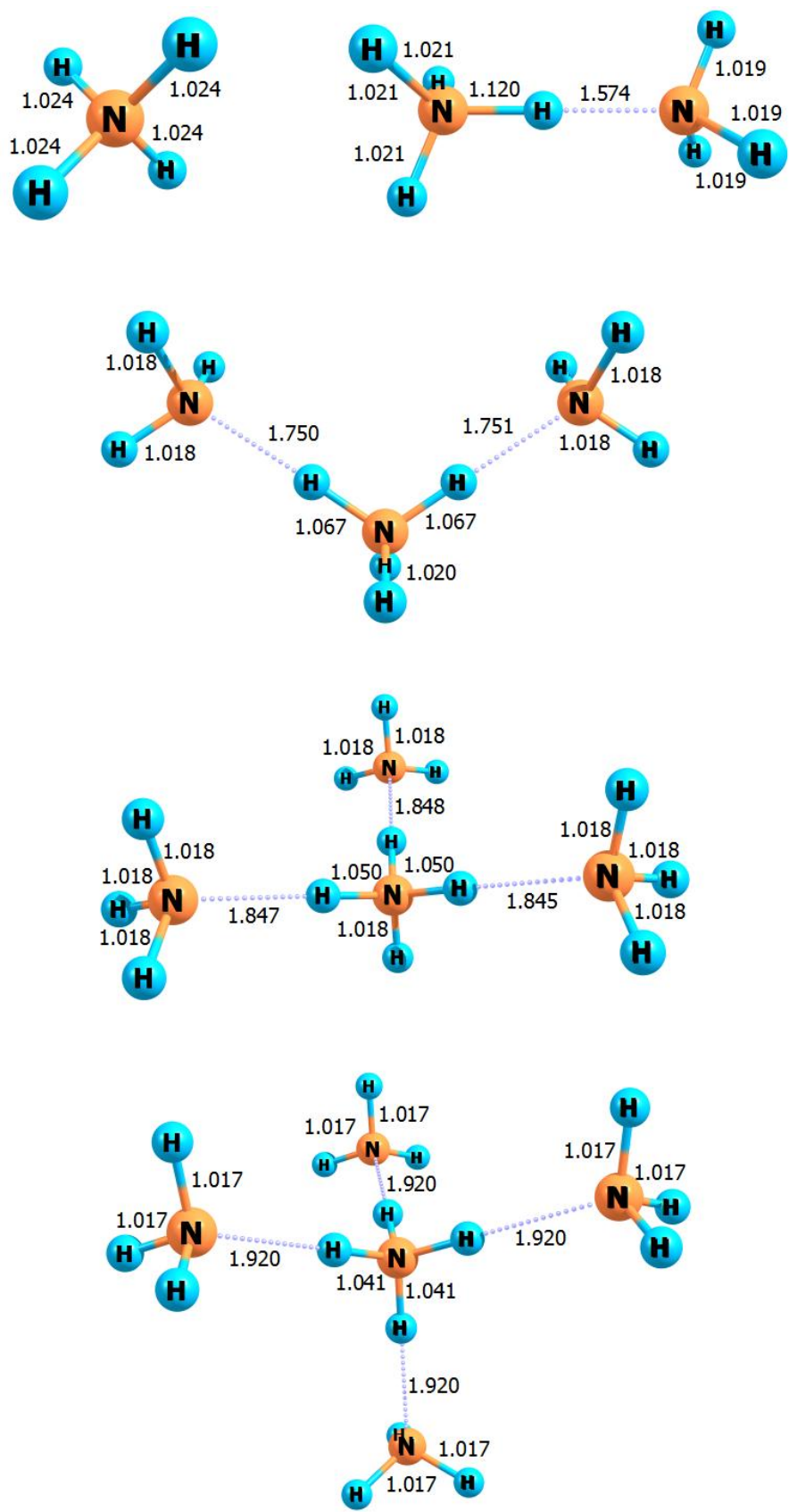

Fig. 1. Equilibrium structures of $N_nH_{3n+1}^+$ cations obtained at MP2/6-311++G(d,p) level. H-bonds are indicated by dotted lines and bond-lengths (in Å) are also given.

## 3. Results and discussion

### 3.1. Structures, bonding and stability of $N_nH_{3n+1}^+$

The structure of $NH_4^+$, with a bond length of 1.024 Å, assumes a tetrahedral ($T_d$) point group of symmetry. In order to obtain the structure of $N_nH_{3n+1}^+$ cations for $n$ = 2–5, we have replaced H-atoms in $NH_4^+$ successively by $NH_4$ units. The equilibrium structures of $N_nH_{3n+1}^+$ cations, thus obtained, are displayed in Fig. 1 along with respective bond lengths. For $N_2H_7^+$, the structure appears as a hydrogen bonded (H-bonded) complex $[NH_4\cdots NH_3]^+$, in which bond-lengths of the $NH_4$ moiety is slightly reduced. Note that $N_2H_7^+$ has been previously studied by several groups.[27-30] The H···N bond distance in this complex is 1.574 Å. This value is in agreement with the previous report of Hou *et al.*[19] and Platts *et al.*[29]. In the case of $N_3H_{10}^+$, there exist two H···N bonds with the bond distance of about 1.750 Å and 1.751 Å. Thus, the structure appears as $[NH_4\cdots 2(NH_3)]^+$ complex in which bond-lengths increase to 1.067 Å. Similarly, the structure of $N_4H_{13}^+$ consists of three H-bonds with the distance ranging 1.845-1.848 Å, and the bond-length of $NH_4$ moiety being 1.050 Å. The structure of $N_5H_{16}^+$, in which all H-atoms are replaced by $NH_4$ moieties, possesses four H-bonds of the distance of 1.920 Å. Thus, the structures of $N_nH_{3n+1}^+$ species are dominated by unusual N-H ···N type of H-bonds, which can be expressed in the form of $[NH_4\cdots (n-1)NH_3]^+$ complexes.

Table 1. QTAIM calculated topological parameters for N-H···N bonds in $N_nH_{3n+1}^+$ cations.

| Cation | No. of H-bonds | $\rho$ (a.u.) | $\nabla^2\rho$ (a.u.) | $V$ (a.u.) | $G$ (a.u.) | $H$ (a.u.) | $\Delta E$ (kcal) |
|---|---|---|---|---|---|---|---|
| $N_2H_7^+$ | 1 | 0.076595 | 0.065522 | -0.077287 | 0.046834 | -0.030453 | 24.3 |
| $N_3H_{10}^+$ | 2 | 0.049719 | 0.091300 | -0.043907 | 0.033366 | -0.010541 | 13.8 |
| $N_4H_{13}^+$ | 3 | 0.039392 | 0.090498 | -0.031699 | 0.027162 | -0.004537 | 10.0 |
| $N_5H_{16}^+$ | 4 | 0.033238 | 0.085818 | -0.024884 | 0.023169 | -0.001715 | 7.8 |

The nature and strength of these H-bonds can be described by the quantum theory of atoms in molecules (QTAIM) analyses. QTAIM reveals the bonding between two atoms by the existence of a bond critical point (BCP) and characterizes it on the basis of topological parameters such as charge density ($\rho$), its Laplacian ($\nabla^2\rho$) and total electronic energy density (*H*) at BCP. According to Rozas *et al.*[31]: (i) $\nabla^2\rho < 0$ and $H < 0$ for strong H-bond of covalent nature, (ii) $\nabla^2\rho > 0$ and $H < 0$ for medium H-bond of partially covalent nature, and (iii) $\nabla^2\rho > 0$ and $H > 0$ for weak H-bond of electrostatic character. The topological parameters at BCP associated with H-bonds in $N_nH_{3n+1}^+$ species are listed in Table 1. It is evident that $\nabla^2\rho > 0$ and $H < 0$ for all N-H···N bonds. Therefore, all H-bonds possess medium strength with partially covalent nature. According to Espinosa *et al.*[32], the interaction energy ($\Delta E$) of H-bonds can be quantified as, $\Delta E = -\frac{1}{2} V$, where *V* is the potential energy density at BCP. The $\Delta E$ of N-H···N bonds in $N_nH_{3n+1}^+$ species are also listed in Table 1. One can note that the H-bond strength in $N_2H_7^+$ is 1.05 eV. Likewise, $N_3H_{10}^+$, $N_4H_{13}^+$, and $N_5H_{16}^+$ possess two, three and four H-bonds with the strengths of 0.60 eV, 0.43 eV and 0.34 eV each, respectively.

Table 2. MP2/6-311++G(d,p) calculated dissociation energy, vertical electron affinity (EA$_v$) and NBO charge on core N-atom ($Q_c$) in $N_nH_{3n+1}^+$ cations.

| Cation | Dissociation paths | Dissociation energy (eV) | EA$_v$ (eV) | $Q_c$ (*e*) |
|---|---|---|---|---|
| $NH_4^+$ | $NH_3 + H^+$ | 9.26 | 4.39 | -0.814 |
| $N_2H_7^+$ | $2NH_3 + H^+$ | 10.45 | 3.52 | -0.884 |
|  | $NH_3 + NH_4^+$ | 1.20 |  |  |
| $N_3H_{10}^+$ | $3NH_3 + H^+$ | 11.34 | 3.00 | -0.916 |
|  | $2NH_3 + NH_4^+$ | 2.08 |  |  |
| $N_4H_{13}^+$ | $4NH_3 + H^+$ | 12.07 | 2.64 | -0.946 |
|  | $3NH_3 + NH_4^+$ | 2.81 |  |  |
| $N_5H_{16}^+$ | $5NH_3 + H^+$ | 12.68 | 2.39 | -0.973 |
|  | $4NH_3 + NH_4^+$ | 3.43 |  |  |

The stability of $N_nH_{3n+1}^+$ species can be verified by analyzing their dissociation into various fragments. We have considered the prominent dissociation paths, such as $N_nH_{3n+1}^+ \rightarrow N_nH_{3n} + H^+$ (deprotonation) and $N_nH_{3n+1}^+ \rightarrow (n-1)NH_3 + NH_4^+$ (loss of ammonia), and calculated dissociation energy as listed in Table 2. The deprotonation of $N_nH_{3n+1}^+$ species is quite difficult due to high dissociation energy against loss of a proton ($H^+$), which further increases with the increase in $n$. Furthermore, the dissociation energies against loss of ammonia ($NH_3$) are very important due to the fact that $N_nH_{3n+1}^+$ cations can be realized as $[NH_4\cdots(NH_3)_{n-1}]^+$ complexes. From Table 2, it is evident that $N_nH_{3n+1}^+$ cations are energetically stable against loss of ammonia as all dissociation energy values are positive. For instance, the dissociation energy of $N_2H_7^+$, i.e., $[NH_4\cdots NH_3]^+$ is 1.20 eV (27.7 kcal/mol), which is in accordance with its H-bond strength (see Table 1) as well as previous estimate at CCSD(T)/6-311++G(3df, 3pd) level.[19] Moreover, the dissociation energy values increase with the increase in $n$. This fact is also consistent with the increase in the number of H-bonds, *i.e.,* net H-bond interaction energies with the increase in $n$.

### 3.2. Superalkali nature of $N_nH_{3n+1}^+$

In order to explore the superalkali nature of $N_nH_{3n+1}^+$ cations, we have computed their vertical electron affinity ($EA_v$) and listed in Table 2. Note that the vertical electron affinities ($EA_v$) of cations reflect the ionization energy (IE) of corresponding neutral species. Evidently, the $EA_v$ of $NH_4^+$ (4.39 eV) is lower than the IE of the Li atom, which is 5.39 eV.[33] Likewise, the $EA_v$ of $N_2H_7^+$ is smaller than the IE of the Cs atom (3.89 eV). With the increase in $n$, the $EA_v$ further decreases monotonically and becomes 2.39 eV for $N_5H_{16}^+$. Therefore, $N_nH_{3n+1}^+$ species form a new series of non-metallic superalkali cations. In order to get some insights into electronic structure, we have observed the highest occupied molecular orbital (HOMO) surfaces of $N_nH_{3n+1}^+$ plotted in Fig. 2. It can be seen that the HOMO of $NH_4^+$ is contributed by the whole molecule. With the increase in the $n$, the HOMO is mainly contributed by $NH_3$ moieties, with a little contribution from $NH_4$ moiety. In $N_5H_{16}^+$, the HOMO is composed of only one $NH_3$ moiety with a marginal contribution from other moieties. However, the HOMOs do not seem to have any direct relation with the $EA_v$ of superalkali cations.

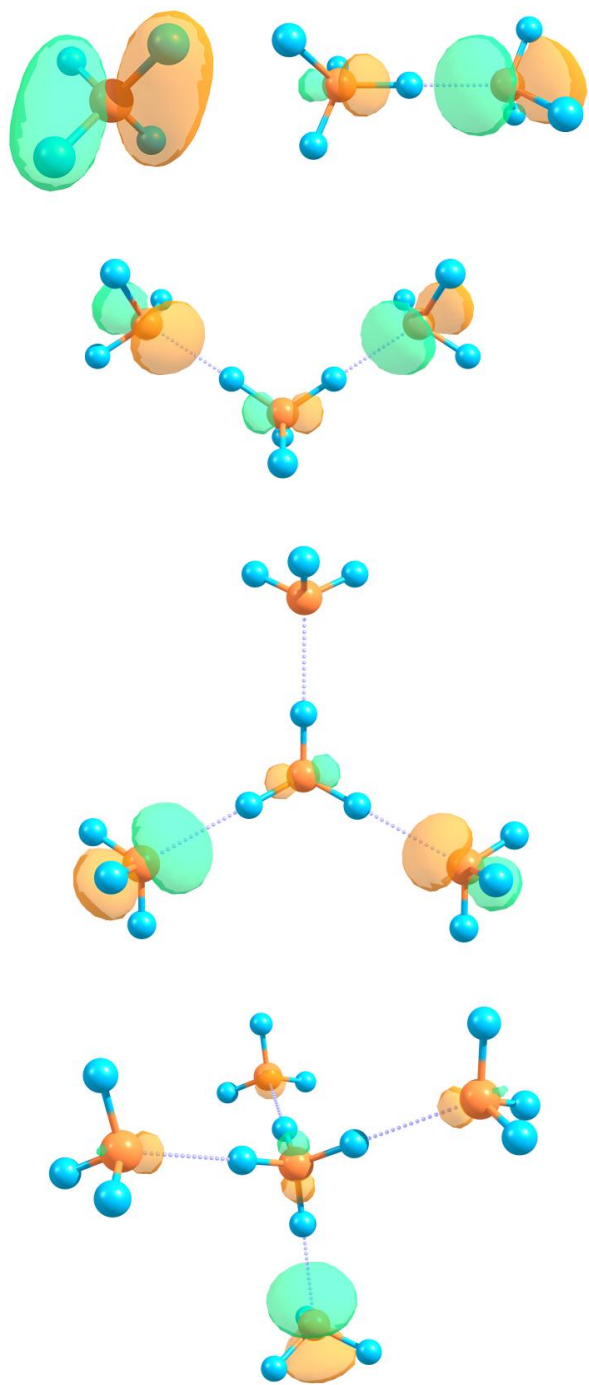

Fig. 2. The highest occupied molecular orbitals (HOMOs) of $N_nH_{3n+1}^+$ cations with an isovalue of 0.01 a.u.

The decrease in the $EA_v$ of $N_nH_{3n+1}^+$ can be explained on the basis of the localization of electronic charge. In Table 2, we have listed the NBO charge localized on central N-atom ($Q_c$) in $NH_4$ moieties. One can note that the localization of the charge increases with the increase in $n$. This may explain the decrease in the $N_nH_{3n+1}^+$ species with the increase in $n$. In Fig. 3, we have shown a correlation graph between $Q_c$ and $EA_v$. One can see that there exists a linear correlation between $Q_c$ and $EA_v$ such that $EA_v = 14.88 + 12.89\ Q_c$, with a correlation coefficient of $R^2 = 0.99412$. Thus, the $EA_v$ of $N_nH_{3n+1}^+$ species shows a good correlation with the $Q_c$ value. This series of superalkalis can be further continued for $n > 5$ by replacing H-ligands of peripheral moieties by $NH_4$. This will add a new member in this family of superalkalis with even lower $EA_v$.

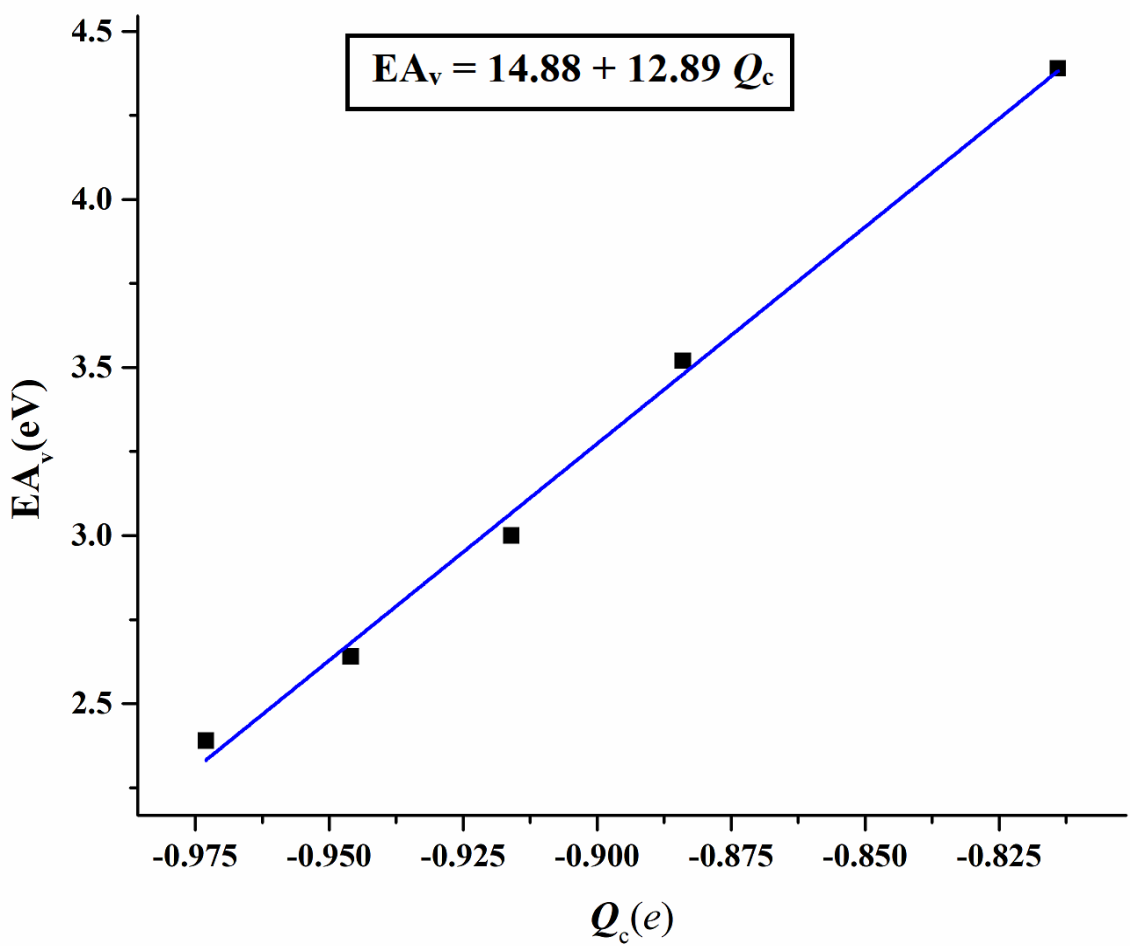

Fig. 3. Correlation between vertical electron affinity ($EA_v$) and NBO charge on core N-atom ($Q_c$) for $N_nH_{3n+1}^+$ cations.

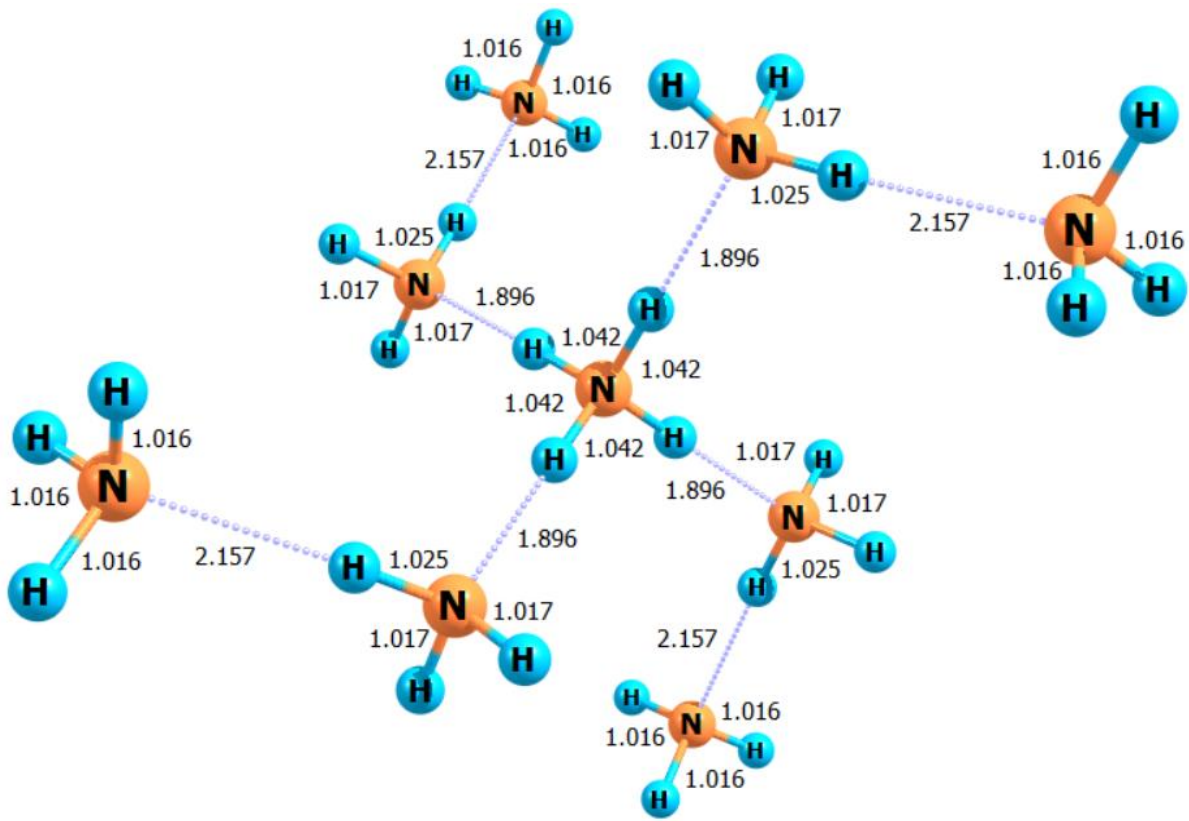

Fig. 4. Equilibrium structure of $N_9H_{28}^+$ cation obtained at MP2/6-311++G(d,p) level.

As mentioned earlier, $N_5H_{16}^+$ was designed by replacing all H-ligands of $NH_4^+$ by $NH_4$ moieties. Now, we replace one H-ligands of these peripheral $NH_4$ moieties by $NH_4$ moieties themselves. This results in an equilibrium structure of $N_9H_{28}^+$ as shown in Fig. 4. One can observe that $N_9H_{28}^+$ cation appears as an $[NH_4\cdots4(NH_3)\cdots4(NH_3)]^+$ complex. Evidently, there exist eight H-bonds, four with the bond-distance of 1.896 Å (HB-1) and four with that of 2.157 Å (HB-2). QTAIM analysis of $N_9H_{28}^+$ reveals $\rho$ = 0.035284 a.u., $\nabla^2\rho$ = 0.088300 a.u., $V$ = -0.027158 a.u., $H$ = -0.002541 a.u. for HB-1 and $\rho$ = 0.019965 a.u., $\nabla^2\rho$ = 0.060114 a.u., $V$ = -0.012200 a.u. and $H$ = 0.001414 a.u. for HB-2. Evidently, HB-1's are partially covalent with medium strength ($\nabla^2\rho > 0$ and $H < 0$) whereas HB-2's are weak with electrostatic nature ($\nabla^2\rho > 0$ and $H > 0$). Likewise, the interaction energies ($\Delta E$) of HB-1 and HB-2 are 8.5 kcal/mol and 0.4 kcal/mol, respectively. Note that the $EA_v$ of $N_9H_{28}^+$ is computed to be 1.84 eV. This value is even lower than half of the IE of Cs atom.

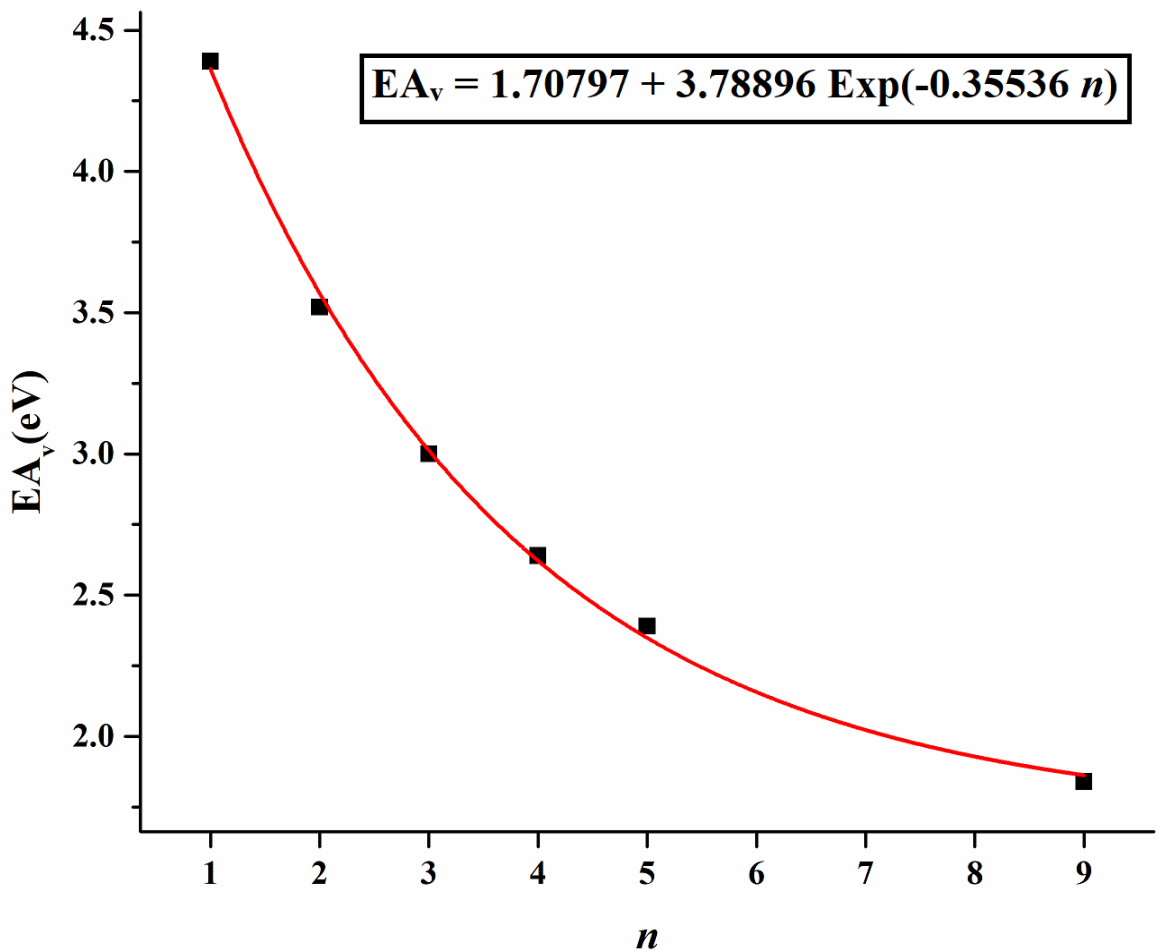

Fig. 5. Correlation between vertical electron affinity ($EA_v$) and number of N-atoms ($n$) for $N_nH_{3n+1}^+$ cations.

Thus, it is indeed possible to continue this series for higher $n$ values by replacing H-ligands successively by $NH_4$ moieties, which will eventually lower the $EA_v$ of $N_nH_{3n+1}^+$ species. This process can be repeated as long as permitted by the repulsion of H-H ligands. Therefore, it seems interesting to analyze the variation of $EA_v$ of $N_nH_{3n+1}^+$ cations with the increase in the number of N-atoms ($n$). In Fig. 5, we have displayed a correlation between $EA_v$ and $n$. There exists an exponential correlation between $EA_v$ and $n$, viz. $EA_v = 1.70797 + 3.78896 \, Exp(-0.35536 \, n)$, with a correlation coefficient of $R^2 = 0.99761$. Statistically, it may not be a good correlation due to a limited number of data points, it does provide a rough estimate of the $EA_v$ of $N_nH_{3n+1}^+$ class of superalkali cations for any desired value of $n$.

## 4. Conclusions

We have studied a new series of $N_nH_{3n+1}^+$ cations and explored their superalkali nature using MP2/6-311++G(d,p) level. The findings of this work are summarized below:

i. $N_nH_{3n+1}^+$ cations possess a number of unusual N-H···N type of H-bonds, which can be expressed in the form of $[NH_4···(n-1)NH_3]^+$ complexes. These H-bonds are partially covalent, whose interacting energy ranges 7.8-24.3 kcal/mol.

ii. $N_nH_{3n+1}^+$ cations are stable against deprotonation $[N_nH_{3n+1}^+ \rightarrow N_nH_{3n} + H^+]$ and loss of ammonia $[N_nH_{3n+1}^+ \rightarrow (n-1)NH_3 + NH_4^+]$.

iii. $N_nH_{3n+1}^+$ species behave as superalkali cations, whose $EA_v$ varies between 4.39 eV to 2.39 eV as $n$ varies from 1 to 5. This can be explained on the basis of electron localization on central core N-atom ($Q_c$), as $EA_v$ correlates linearly with $Q_c$.

iv. $N_nH_{3n+1}^+$ series can be continued to include new superalkalis with even lower $EA_v$ values. For instance, the $EA_v$ of $N_9H_{28}^+$ cation is computed to be 1.84 eV.

v. There exists an exponential relation between $EA_v$ of $N_nH_{3n+1}^+$ series with the number of N-atoms ($n$), which may be useful in predicting $EA_v$ for any desired value of $n$.

**Acknowledgement**

Dr. A. K. Srivastava acknowledges Prof. N. Misra, Department of Physics, University of Lucknow, for providing computational support and Prof. S. N. Tiwari, Department of Physics, DDU Gorakhpur University, for helpful discussions.